# How managing more efficiently substances in the design process of industrial products? An example from the aeronautics sector

Lemagnen, M.[1,2], Mathieux, F.[1], Brissaud, D.[1],
1 Laboratoire G-SCOP, fabrice.mathieux@g-scop.inpg.fr, daniel.brissaud@inpg.fr;
2 Hispano-Suiza, maud.lemagnen@hispano-suiza-sa.com

**ABSTRACT**

As environmental concerns increase through states, industries and worldwide organizations, companies have to implement these new requirements in their different activities. Lowering environmental impacts of products, i.e. ecodesign, is considered today as a new and promising approach environment protection. This article focuses on ecodesign in the aeronautical sector through the analysis of the practices of a company that designs and produces engine equipments. Noise, gas emissions, fuel consumptions are the main environmental aspects which are targeted by aeronautics. From now on, chemical risk linked to the use of materials and production processes has to be traced, not only because of regulation pressure (e.g. REACh) but also because of customers requirements. So far, the aeronautical sector hasn't been focusing much on managing chemical risks at the design stage. However, new substances regulations notably require that chemical risk management should be by industries used as early as possible in their product development process. In order to comply with the latter, the aeronautics sector has to elaborate chemical risk management tools coping with complex design and numerous data requirements.

The aim of this paper is to present a new method hat should be adapted to aeronautical designers' practices and based on simple entry data, but efficient enough to ensure a good substances traceability all along the product design process. The method second objective is also to guide the designer's materials and process choices by avoiding the most chemically critical ones. The proposed method has been elaborated by an industrial dedicated team and is based on four main steps. The structure, the related tools and the parameters of the method have been chosen and developed in order to be easily understood by non environmental experts, and directly operational.

The paper reports also the test of the method on an engine equipment, the ECU (electronic control unit), the test permitted to validate the relevance of the method to design activities. It also underlined aspects of the method to be detailed and developed, such as the parameters definition. Further works are also described.

## I. INTRODUCTION: HELPING COMPANIES TO BETTER MANAGE SUBSTANCES

### 1. Context

As damages to the environment, related to plant activities and effluents, are nowadays quite well controlled through a set of regulation texts and clean technologies, those related to products are barely covered. However, the link between products and potential hazard due to their composition has increasingly been noticed for the past decade. At the European level, three past Directives started to focus on particular issues among environmental impacts generated by products. These texts, which are applied to three different families of products (Pack: packaging [1], RoHS: electronic equipments [2] and ELV: cars [3]), have in common to limit the use of some chemical substances (e.g. Pb, Hg, Cd) into the products. Pack, RoHS and ELV notably aim at reducing the toxicity risk at the end of life of the products.

Solutions had to be found to adapt industrial production activities to these new requirements. Examples of such adaptations can be illustrated by electronic and automotive sectors, through technologies evolution (lead-free brazing processes development) or substances management all along the supply chain (through for example the IMDS system).

REACh [4], a European regulation adopted in 2006, gives new orientations to the European substances-orientated policy. Three aspects of REACh are of key importance:
- There should have complex mechanisms of registration and evaluation of chemical substances put in the market, which tends at last to eliminate the most toxic ones.
- The products' chemical composition should be known through masses parameters.
- The potential toxic impact of products should be determined during its life-cycle.

REACh does not only require that industrial companies control their supply chains, but also that they are able to quantify the amount of toxic substance at each step of the product life. Chemical substance toxicity should therefore be taken into account by designers as a new criterion, together with any other environmental impact categories. This is in line with current LCA method development [5].

### 2. Objectives

As suggested initially by Safran group [6], REACh brings a new complexity to industrial activities as companies should comply with its requirements from the early steps of product development. We think at Hispano-Suiza that one of the key answers to REACh compliance may lie in the capacity of carrying out a process of chemical risk assessment for substances, materials and production process at any step of the design. As recommended by [7], integrating environmental aspects during the design process means that an environmental assessment should be lead as a baseline on a reference product. Environmental impacts can be then quantified and classified into categories, and even prioritized. This approach would help the designers to choose the best solutions with an aim of decreasing environmental burden of the product life-cycle. The results of the assessment should be presented so that designers can operate a choice between, substances, mixtures, materials or production process [8].

This article presents a new method developed for the aeronautics sector that should help design decision making through chemical risk quantification. Section 2 introduces the method itself while Section 3 reports its application to an aeronautical product. Last Sections focus on discussion, conclusions and perspectives.

## II. PRESENTATION OF A NEW METHOD TO BETTER MANAGE SUBSTANCES

### 1. Industrial context of the method development

The proposed method is developed in an aeronautical company, Hispano-Suiza (Safran group). Hispano-Suiza designs and produces aeronautical engine equipments and engine regulation systems. Probably because noise and gases emissions are considered to be the main environmental impacts generated by engines, aeronautics sector is however still at the early stages of taking a multicriteria approach on environmental product management, especially in comparison with automotive or electronic sectors. Furthermore, until recently, aeronautics has not been much concerned with substances regulation, mainly because of flight safety requirements and of related exemptions. Because of the large scope of new European directives (RoHS, REACh), aeronautical industries begin to be impacted by substances requirements, in general indirectly through suppliers. Direct impacts can however not be ignored in short-term.

One of the first initiatives of chemicals traceability lays in the construction of substances' lists, defined through their CAS number, their risk-phrase and their legal status: "banned" / "exempted for particular use" / "authorized but critical". Several lists have been proposed since 2000 by Safran (ex-Snecma), Airbus and DGA (French General Direction of the Army). Some of these lists have then been retrieved in 2007 by the aeronautical and spatial industries French Federation (GIFAS), in order to propose a common list [6]. Lists have often been used as an argument toward ISO 14001 environmental auditors to illustrate the product-centered environmental approach (forbidden materials, materials specification from ordering party to their article suppliers). Even European aeronautics sector uses this type of list as one step to take into account REACh regulation requirement. However, it does not seem to entirely comply with the whole regulation requirements.

### 2. Specification of the method

As noticed above, the aeronautics sector seems to lack tools to efficiently manage toxicity-related issues. There is indeed a need to develop a method to do so. This method should in particular meet the three following objectives:

- lead to results that are usable by designers to orientate their choices: indeed, expertise on toxicity issues is usually the field of small communities, such as risk assessment public institutions. To be efficient from a design perspective, the method should not require any particular knowledge on toxicity, so that it can be used by designers. Moreover, results should be simple to interpret by toxicology non - experts [9]; they should also properly discriminate several elements, from non hazardous to hazardous.

- be implemented with data available in engineering departments [10]: data collected from any accurate design documentation should be sufficient to be used as entry data : material type, mixture commercial name, production process specification reference. Method should be considered by the designer as a "blackbox" that doesn't necessitate any intermediate calculation to be run;

- answer REACh compliance approach defined in Section 1.1: in particular, toxicity level knowledge isn't accurate enough to answer REACh requirements. Substances amount data should also be available, in order to evaluate the product compliance with very hazardous substances thresholds.

### 3. Global structure of the method

As toxicity is considered as an environmental impact, its treatment trough the design process could be made through:

- Hazard assessment of substance, mixtures, materials and production processes,
- Evaluation of the amount of toxic substances contained in mixtures, materials and production processes.

In order to comply with these requirements, the proposed method is composed of 4 steps, described below from Step 1 to Step 4

#### a. Step 1: Technical data extraction step

The necessary data to be used in the method concern all the materials, mixtures and production processes which are used to produce the article[1]. The data is usually found in the products' bill of materials or can be extracted from product management information systems. However, not all accurate data can be found, because sometimes an element is chosen for its function, and not for its composition.

#### b. Step 2: Hazard qualification step

1. Description

As the designer possesses the data needed to run the method describing the materials, mixtures and production process that make the product, he proceeds to the step of their labelisation. We suggest using a basis criterion that is the chemical substances list [6], built according to three levels. Materials, mixtures and production processes are also organized into the same three levels, according to the substances they are composed or realized with. We decided of some rules, which are very simple, to pass from substance to materials, mixture or production processes. The rules, which are not described in this paper, have notably been elaborated using the Directives 1999/45/EEC and 67/548/EEC. Figure 1 illustrates the link between substance and materials or mixtures labelisation.

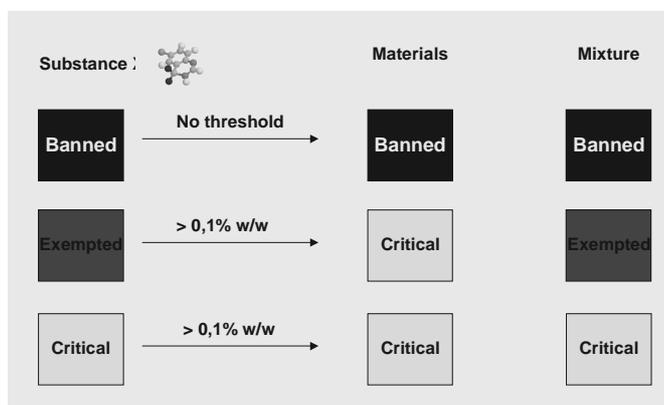

*Figure 1: Labelisation rules from substance to materials and mixture*

When any substance is under 0,1% w/w threshold, materials and mixtures are considered free labeled.

2. Database

In order to compile the label results for all materials, mixtures and production process of the product, a database as been built by Hispano-Suiza [11].In a primary approach, it compiles all the different materials, mixtures and production process used specified by the company. That database is firstly bound for designers and will quicken further requests for new complete assessments. The database displays elements such as substance, materials, mixture or production process label, but also others data which have been chosen because of their relevance for the user (commercial mixture name or a design specification reference).

---

[1] "Article" stands for any object which shape and design is more relevant for its function than its chemical composition [REACh]

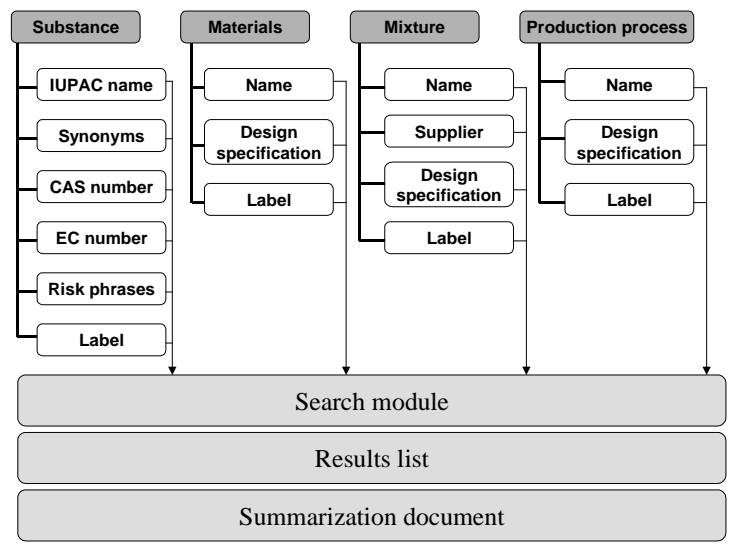

*Figure 2: Database structure*

c. Hazard quantification step

The database presented above provides a function that permits to compile all the materials, mixtures or production processes that are requested for a product development. It is possible to create a figure (indicator) which summarizes the number of each materials, mixture or production process according to its label. However, at this stage of the method application, one can find numerous materials under the same label. In order to discriminate the results presented above, it was decided to use a quantitative chemical risk assessment method for exempted and critical materials, mixtures and production processes. This two steps (qualification + quantification) approach helps to filter the data and concentrate on the most critical identified values, as usually requested by traditional H&SE practitioners (cf. e.g. [12]).

The method which is employed is inspirited from the French INRS chemical assessment risk methodology presented in [13]. This method has been chosen because of its ergonomics: indeed, only few parameters are needed as input; furthermore, these parameters are intuitive for any people in an industrial environment. This method is originally used in the context of production plant management in Small and Medium size Enterprises (SME), in order to help H&SE engineers to manage the chemical products risks (use, storage). Figure 3 indicates the structure of the method.

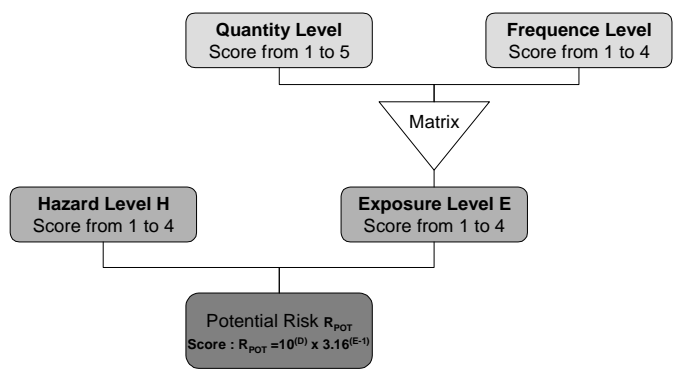

*Figure 3: INRS chemical risk assessment method structure (adapted from [13]).*

We adapted some of the parameters with the data extracted from the product design bill and determinate correspondences between the INRS scores leveling. In order to comply with the calculation system of the initial method, we adapt danger, quantity and exposure parameters to the same 4 or 5 levels scales used. We verified the correct equivalence between INRS method and the scale we chose.

d. Decision making step

The quantification step is crucial to guide the designer's team environmental choice. The INRS method suggests a way to prioritize the potential risk through the setting of thresholds for the scores obtained. According to [13], three levels have been defined (<100; 100 < … < 1000; > 1000), and for each of them, an orientation of action is delivered, as presented in the following figure. This approach was chosen for our method.

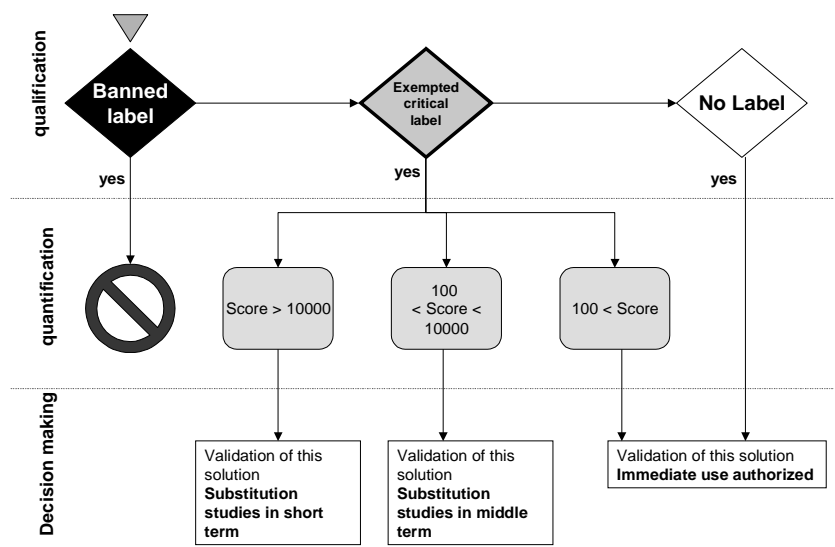

*Figure 4: Decision making guidance according to qualification and quantification assessment results*

In order to file the assessment results and to justify the technical choices made during the design process, we completed the approach with a technical document called "compliance matrix". This document is simultaneously filled in as the assessment is conducted. The matrix is today required as any technical document during the design reviews and is examined.

e. Summary of the method

The method is summarized in figure 5:

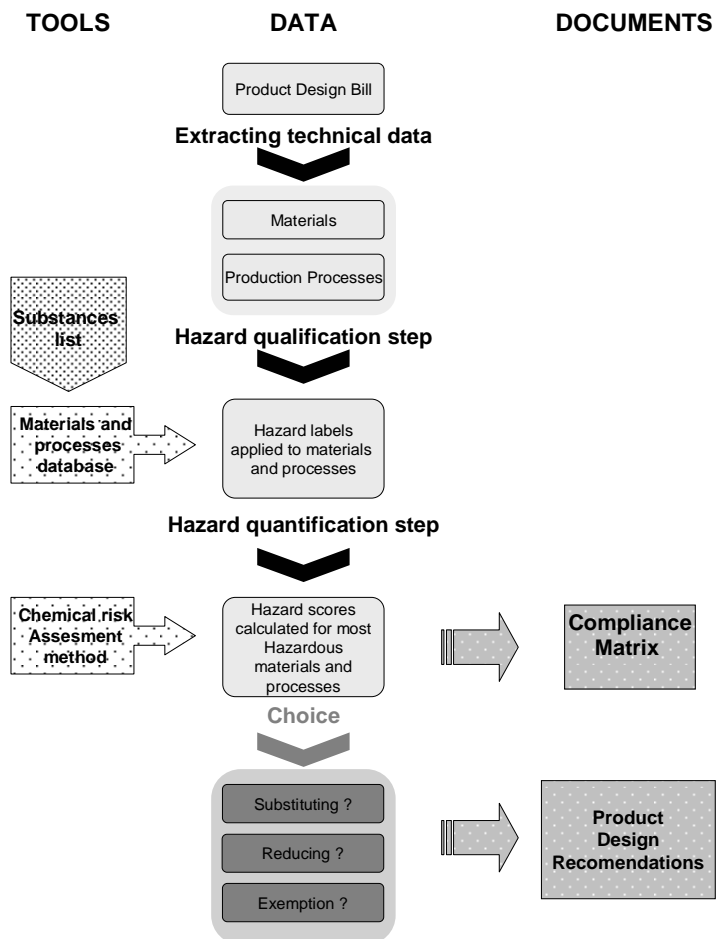

*Figure 5: Overview of the method steps.*

## III. CASE STUDY: APPLYING THE METHOD TO ELECTRONIC CONTROL UNIT FOR TURBOSHAFTS ENGINE

The method presented above has been applied on aeronautical equipment (cf. Figure 6), designed and produced by Hispano-Suiza ([6], [11]). The equipment is an Electronic Control Unit (ECU) for turboshafts engine for both civil and military helicopters. This equipment is connected to the engine and its main functions are to regulate the engine and to ensure its functional safety. Basically, the ECU is an intermediate between the pilot and the engine, and runs according to the 4 following operations:

- gathering data on engine, sent by sensors via wiring harnesses,
- filtering data, due to severe environment (electromagnetic interferences …),
- analyzing data and calculating commands to be sent to the control devices
- formatting commands to drive the actuators and other control devices.

The equipment was chosen because it is representative of the main technologies (materials and production processes) used at Hispano-Suiza. The parts of this equipment are:

- mechanical part: housing, armour plating, screws, various metallic parts which represent about 45% of total ECU mass,

- electric/ electronic parts: 4 electronic printed wired boards, internal and external connectors (around 55% of the ECU mass).

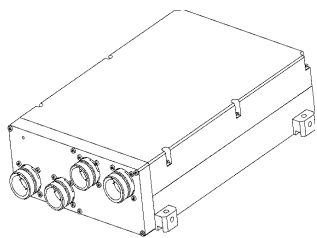

*Figure 6: An Electronic Control Unit.*

As numerous aeronautical equipments, which life lasts around twenty five years and due to the technical evolutions or customers requirements, the ECU is subjected to regular redesign processes. The method has been applied during the redesign phase of the product. Most of the Technical data have been extracted from the design specifications. The experience showed that some data was not directly available from the design plans, and calculations as well as measurements on previous equipment generation were necessary. As a first approximation and because of time constraints the assessment was made without considering the electronic components. However, all electronic components used are certified as RoHS compliant, critical chemicals should not be found in these parts. The assessment leads to the following results, represented on figure 7:

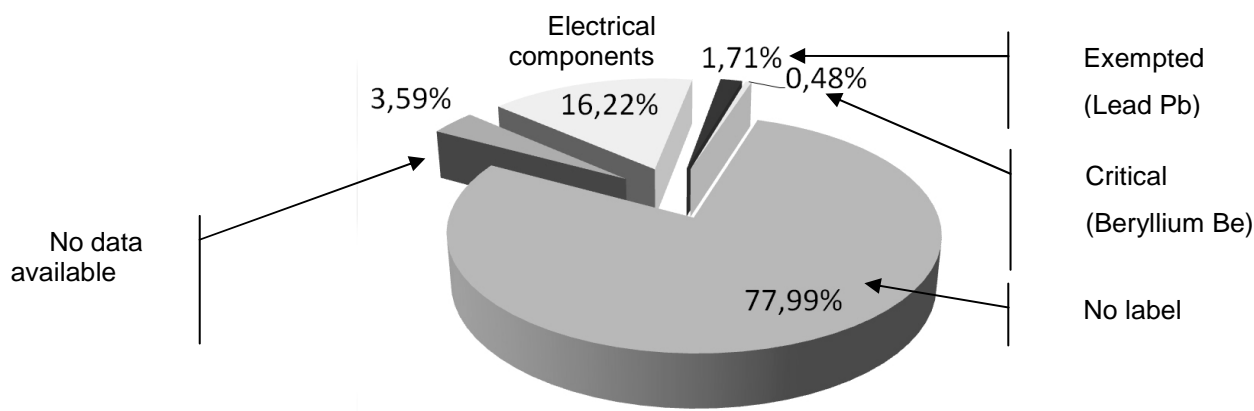

*Figure 7: Assessment test results – Materials label repartition quoted on total ECU mass*

After analyzing Figure 7, it can be concluded that:

- Most of the ECU parts do not present any toxicological or chemical risk, according to the substances list standard,
- About 3,6 % of the ECU mass could not be assessed, because of lack of data. This percentage corresponds to furnished articles, which are bought to answer technical requirements and function, and which structure is unknown for the customer (i.e. Hispano-Suiza), because of industrial property.
- Two hazardous substances concentrations are higher than 0,1% weight/weight : lead in brazing and beryllium in a specific alloy. These substances should be notified to the customers in order to comply with REACh requirements.

From this assessment, the design team took action to improve data collection: a questionnaire has been created to be distributed to suppliers, for furnished articles. So far, a few REACh conscious suppliers have already answered to our questionnaire.

The following Design recommendations have been notified by the design team in the documentation which accompanies the equipment:

- Lead substitution: lead brazing processes are still used to produce the ECU. Even though free-lead brazing processes haven't been totally validated, aeronautical Research and Development activities still cover that field. We hope that the a brazing process change will be possible for the next generation of ECU,
- Beryllium alloy risks management: this alloys presents particular elasticity properties, but its machining produces carcinogenic dust. Thanks to the current state of knowledge, no other material could perform these properties. In order to limit risks during processing phase, we explicitly require from our suppliers to reduce exposure factors.

## IV. DISCUSSION

Thanks to the case study analysis, we can conclude that the method proposed in this paper seems to be:

- applicable during design as it uses information on materials, mixtures and production processes available at design stages;
- usable by designers as assessment results are precise enough to guide designer's choice between different elements;
- Efficient to answer REACh requirements about substances mass thresholds: the method results point out the most hazardous substances mass ratio in the equipment.

The method presents also some limits, in particular:

- The exposure step calculation has mainly been applied to the production and maintenance phased of the equipment: this should be extend in the future to other life-cycle stages,
- the sources of input parameter should be explored further , so that uncertainties linked to measurement on former generation equipment are reduced.

## V. CONCLUSION / PERSPECTIVES

The present paper aimed at presenting a new chemical risk assessment method, to be used during the design process of a product. This method has been (and is still being) developed in an aeronautical company, in order to answer its particular requirements. We however think that this method could be of use for other companies of the sector. The method is a combination of a qualitative and quantitative approach. The main elements are:
- a substances list, which is a classification of various substances according to their toxicological and chemical intrinsic hazard,
- a database which connects substances list to operational data used by designers: materials, mixtures and production processes,
- a risk assessment which permits to guide the designer's choice among different materials, mixture or production processes, thanks to their environmental weight.

The application of the method on a case study has been lead and conclusions have been made:
- entry data, necessary to run the method, have been validated,
- the assessment results turns possible to formulate design recommendations, and to justify from substitution research orientation,
- the assessment results permit to answer the REACh requirement concerning hazardous substance notification in products to be sold.

Further works includes:

- refining the input parameters used in the method
- testing the method in other design projects activity in the company,
- extend the assessment on electronic components
- Exposure assessment should be refined

**ACKNOWLEDGEMENT:**

The authors would like to thank L.-F LOBREAU, R.POIRIER, S.COCHARD, C.GALLAIS, P.E. PLACHOT, P.BERHAULT and Q.MOLLET for their involvement.